\documentclass[a4paper,11pt,english]{elsarticle} 
\usepackage[utf8x,utf8]{inputenc}
\usepackage{lineno,hyperref}
\usepackage{color}
\hypersetup{colorlinks = true, linkcolor = red, citecolor = blue, bookmarksopen = true}
\usepackage{caption}
\usepackage{amsmath, dsfont} 
\usepackage{amsthm}
\usepackage[left=3cm,right=3cm,top=3cm,bottom=3.5cm]{geometry}
\usepackage{stmaryrd}
\usepackage{bbold} 
\usepackage{bm}
\usepackage{epstopdf}
\usepackage[normalem]{ulem}

\usepackage{booktabs}
\usepackage{array}

\usepackage{graphicx}
\usepackage{caption, subcaption}

\usepackage[dvipsnames,svgnames,x11names]{xcolor}

\usepackage{calc}
\usepackage{ulem}
\usepackage[ruled,linesnumbered]{algorithm2e}

\usepackage{lmodern}
\usepackage{babel}
\usepackage{enumitem}
\usepackage[all]{xy}
\usepackage{array,multirow}
\usepackage{pgfplots}
\usepackage{tikz,tkz-tab}
\usetikzlibrary{patterns}

\usepackage{multicol}
\usepackage{titletoc}
\usepackage{setspace}
\usepackage{hyperref}

\theoremstyle{definition}

\newtheorem{rem}{Remark}

\newcommand{\RR}{\mathds{R}}
\newcommand{\PP}{\mathds{P}}
\newcommand{\EE}{\mathds{E}}
\newcommand{\NO}{\mathcal{N}}
\newcommand{\II}{\mathds{I}}

\newcommand{\correction}[1]{\textcolor{black}{#1}}

\begin{document}

\begin{frontmatter}

\title{Improvement of the cross-entropy method in high dimension \correction{for failure probability estimation} through a one-dimensional projection without gradient estimation}

\author[a,b]{Maxime El Masri}
\ead{melmasri@onera.fr}

\author[a]{J\'er\^ome Morio}
\ead{jerome.morio@onera.fr}

\author[b]{Florian Simatos}
\ead{florian.simatos@isae.fr}

\address[a]{ONERA/DTIS, Universit\'e de Toulouse, F-31055 Toulouse, France}
\address[b]{ISAE-SUPAERO and Universit\'e de Toulouse, Toulouse, France}

\begin{abstract}
   Rare event probability estimation is an important topic in reliability analysis. Stochastic methods, such as importance sampling, have been developed to estimate such probabilities but they often fail in high dimension. In this paper, we propose a new cross-entropy-based importance sampling algorithm to improve rare event probability estimation in high dimension. We focus on the cross-entropy method with Gaussian auxiliary distributions and we suggest to update the Gaussian covariance matrix only in a one-dimensional subspace. For that purpose, the main idea is to consider the projection in the one-dimensional subspace spanned by the sample mean vector, which gives an influential direction for the variance estimation. This approach does not require any additional simulation budget compared to the basic cross-entropy algorithm and we show on different numerical test cases that it greatly improves its performance in high dimension.
\end{abstract}

\end{frontmatter}

 \section{Introduction}
 
\noindent \textit{Classical algorithms for reliability analysis.} In reliability analysis, an important goal is to determine the probability of failure of an engineering system, often represented as the probability that a function, called limit-state or \correction{performance} 
function, exceeds a given threshold. \correction{Estimating directly this probability can be difficult} 
because calls to the function can be computationally expensive and the failure of the system is often a rare event.
 
\correction{Various techniques have been developed} to estimate such rare event probabilities~\cite{bourinet2018reliability, morio_survey_2014}. First/second order reliability methods (FORM/SORM~\cite{der_kiureghian_first-and_2005}) seek the most probable failure point (or design point) which minimizes the \correction{distance, in the standard normal space,} between the failure hypersurface and the origin, and then realize a linear (FORM) or second-order (SORM) approximation of the limit-state function at this design point in order to estimate the probability with this approximation. Sampling techniques based on Monte Carlo simulations take advantage of the law of large numbers for rare event probability estimations~\cite{rubinstein_simulation_2017}, where this probability is estimated as the average number of random samples which fall in the failure domain. Another sampling alternative is the subset simulation method~\correction{\cite{au_estimation_2001,song2009subset,jiang2021recursive}}, or adaptive splitting, that expresses the probability as the product of several conditional probabilities.

In this article, we focus on the Importance Sampling (IS) method~\cite{owen_safe_2000, agapiou_importance_2017} that is another well-known Monte Carlo method considered in rare event simulation. IS consists in sampling from an alternative density, called auxiliary density, for which the failure is more frequent. Different techniques exist to find an efficient auxiliary density, as this choice has a strong influence on the performance of the overall IS estimation scheme. \correction{The performance comparison of IS with other failure probability estimation techniques is beyond the scope of this article. However, some insights are proposed in the recent article \cite{moustapha2021generalized} that analyses the results of classical Monte Carlo, subset simulation and importance sampling, combined with a surrogate model on different reliability problems.} One of the most well-known method is the cross-entropy method \cite{rubinstein_cross-entropy_2011, de_boer_tutorial_2005}, which we will simply refer to as CE in the rest of the paper. CE aims to determine the best auxiliary density within a given, usually parametric, family of densities: here ``best'' refers to the density which minimizes the Kullback--Leibler (KL) divergence with the optimal auxiliary density, which is intractable as it involves the unknown probability. \correction{Many papers are still published on the CE method, aiming at improving or applying CE in various contexts (\cite{papaioannou_improved_2019,cao2019cross,chan_improved_2012,mattrand2014cross}).} When the goal is to estimate a probability, a variant of CE (called improved cross-entropy method and referred to as iCE in the sequel) has also been developed in~\cite{papaioannou_improved_2019} to improve its efficiency by using a smooth approximation of the indicator function of the failure event. CE and iCE, which are the basis of the present work, are recalled in the Appendix. CE belongs to a larger family of sampling algorithms called Adaptive Importance Sampling (AIS) \cite{bugallo_adaptive_2017,portier2018asymptotic}, where one seeks for a valuable auxiliary density in a parametric family by successive updates of the parameters. \correction{A non-parametric IS has also been proposed, considering kernel density estimation to approximate iteratively the optimal auxiliary density (see \cite{morio2011non} or \cite{neddermeyer2009computationally}).}
	\\ \correction{
		\noindent \textit{New Challenges in importance sampling for reliability analysis.} 
		While the fundamentals of IS are relatively well established, IS suffers from several limitations that may hinder its applicability. First of all, IS often requires a computational budget of at least a thousand calls to the simulation code, which can be too prohibitive in practice. To this end, it is interesting to couple surrogate models and IS and different approaches have been described to optimally achieve this coupling \cite{xiao2020reliability,zhang2020akois,zuniga2021structural}. The enrichment of the surrogate model and particularly the choice of the calculations \cite{yang2020system} to be carried out or the characterisation of the uncertainty \cite{MENZ2021102116} due to the surrogate model on the probability estimation are among others emerging research questions. Unlike classical Monte Carlo approaches, IS is not immediately suitable for high-dimensional problems because the search space for an efficient auxiliary density becomes very large \cite{neddermeyer2009computationally,katafygiotis_geometric_2008,au_important_2003} and also because of the degeneracy of the IS weights \cite{Bengtsson08:0}. To counter this limitation, new approaches for intrinsic dimension analysis and dimension reduction \cite{zahm2018certified,
  uribe_cross-entropy-based_2020} have recently been proposed to focus on the dimensions most influential on integration. We give more details about this issue on the specific case of CE in the next paragraph.
}
		\\
	\noindent \textit{Rare event probability estimation with CE in high dimension.} All these techniques face the curse of dimensionality and become inaccurate as the dimension increases. 
	\correction{For CE this is due to the weight degeneracy phenomenon in high dimension, which involves the collapse of the estimated covariance matrix to $0$ due to weight degeneracy~\cite{Bengtsson08:0, El-Laham18:0}}. In~\cite{rubinstein_how_2009}, auxiliary densities focus on the most influential variables which are selected beforehand, to reduce the number of updated parameters and decrease the variance of the IS estimator. However the technique is not efficient when all variables are influential.~\cite{chan_improved_2012} proposes to directly generate a sample from the optimal unknown IS distribution with an MCMC method and then estimate the CE-optimal parameter. This one should be more accurate and thus lead to a better probability estimation, but in high dimension the MCMC step can require a large simulation budget to reach convergence. Another idea to improve CE in high dimension is to use different, lower-dimensional parametric families of auxiliary densities. For instance,~\cite{wang_cross-entropy-based_2016} propose to use von Mises--Fisher mixtures and~\cite{papaioannou_improved_2019} to use von Mises--Fisher--Nakagami mixtures: both distributions are supported by the hypersphere which reduces the number of parameters compared to Gaussian mixtures~\cite{geyer_cross_2019} which are not efficient in high dimension. These methods are mainly well suited to multimodal problems, that will not be tackled in this paper, and require a large number of samples.  
	
	Closer to the spirit of our paper, the recent paper~\cite{uribe_cross-entropy-based_2020} proposes to construct a subspace, called the Failure-Informed Subspace (FIS), which identifies a low-dimensional structure of the problem and then applies CE on the FIS by projecting the samples: the resulting algorithm is a variation of iCE called iCEred. The FIS is chosen based on a theoretical bound which suggests that it induces a small KL divergence: however its computation requires the evaluation of the gradient of the limit-state function. This presents two drawbacks: 1) the 
	\correction{limit-state} function needs to be differentiable and 2) computing its gradient is often expensive (and sometimes even out of reach).
	\\
	
	\noindent \textit{\correction{Novel methodological contribution.}} In the present paper, we propose two new algorithms, CE-$m^*$ and iCE-$m^*$, which pursue the line of research recently initiated in~\cite{uribe_cross-entropy-based_2020}, namely circumventing the curse of dimensionality faced by CE (and iCE) \correction{for the estimation of a failure probability. The main idea of this article is to project the generated samples on a suitable low-dimensional space. The proposed projection direction is the sample mean vector of the input conditioned on failure (see Section~\ref{sub:CE-m} for more details) whose computation does not require any additional model evaluations compared to CE or iCE as it is not based on the gradient}. This direction is motivated by heuristic arguments presented in Section~\ref{sub:justification} and on all the numerical examples that we considered, in particular the representative ones that we report in Section~\ref{SectionRes}, \correction{they improve the performance of CE.} In many cases and even for simplistic linear 
	\correction{performance} functions, CE and iCE do not converge for dimensions larger than $30$ whereas our algorithms provide accurate estimates up to dimensions $200$ and above. 
	\\

	\noindent \textit{Organization of the paper.} The following study is divided in four sections. Section~\ref{SectionIS} recalls the foundations of IS, CE and iCE. Section~\ref{sec:CE-m} explains why CE does not work in high dimension and then introduces the two proposed algorithms, CE-$m^*$ and iCE-$m^*$. Their performance is numerically evaluated in Section~\ref{SectionRes} on four different examples. We conclude the paper in Section~\ref{Concl}, giving a short summary and some perspectives.
	
 \section{The cross-entropy method}\label{SectionIS}
	
	\subsection{Rare event probability estimation} \label{sub:rare}
	Let $X$ be an $n$-dimensional random vector from a continuous probability distribution function $f$ modeling the input variables. Define a failure domain $\mathcal{F} =\{x\in\RR^n: \varphi(x)\geq 0\}$ and the associated probability of failure as: 
	
	\[ P=\PP_f(\varphi(X)\geq 0)=\EE_f(\II_{\{\varphi(X)\geq 0\}})=\int_{\RR^n} \II_{\{\varphi(x)\geq 0\}}f(x)\textrm{d} x \]
	where $\II_{\{\varphi(x)\geq 0\}}=1$ when $\varphi(x)\geq 0$ and $0$ otherwise. Throughout the paper $\PP_f$ and $\EE_f$ denote the probability and the expectation with respect to $f$. The function $\varphi$ represents a computationally demanding engineering model and is considered as a black box, so that the probability $P$ cannot be estimated analytically. Naive Monte Carlo methods with independent and identically distributed (i.i.d.) samples from $f$ can be used to estimate~$P$, but they require a large number of samples when the failure is a rare event.
	
	\subsection{Importance sampling (IS)}\label{sub:IS}
	
	IS \cite{owen_safe_2000,agapiou_importance_2017} consists in sampling the input variables from an alternative distribution $g$ 
	\correction{well suited} to the failure domain, that is, such that the probability of failure with respect to $g$ is not too rare. Let $g$ be a probability density function such that $g(x)=0 \Rightarrow \II_{\{\varphi(x)\geq 0\}}f(x)=0$, then we have 
	\[ P=\int_{\RR^n}\II_{\{\varphi(x)\geq 0\}}L(x)g(x)\textrm{d}x=\EE_g \left( \II_{\{\varphi(X)\geq 0\}}L(X) \right) \]
	where $L = f/g$ is called the likelihood ratio (or importance weight), and $g$ is the auxiliary density. The IS estimator of $P$ is then
	\[ \hat P=\dfrac{1}{N}\sum_{i=1}^N \II_{\{\varphi(X_i)\geq 0\}}L(X_i) \]
	where $X_1,\ldots,X_N$ are i.i.d. random vectors drawn from $g$. The auxiliary density has a strong influence on the accuracy of the estimation. The optimal auxiliary density $g^*$ is the distribution conditioned on the failure domain, i.e.,
	\begin{equation}\label{gopt}
		g^*(x)=\dfrac{\II_{\{\varphi(x)\geq 0\}}f(x)}{P}, \ x \in \RR^n.
	\end{equation}
	With this distribution, the probability estimate $\hat P$ is exactly $P$ and has zero variance. But $g^*$ depends on $P$, the unknown probability, hence it cannot be used in practice. For this reason, one often looks for an approximation of $g^*$ in a parametric family of densities $\mathcal{G} = \{g_\lambda : \lambda\in\Lambda \}$ called auxiliary density family. A natural choice for $\lambda$ is the parameter $\lambda^*$ that minimizes the variance, i.e.,
	\[ \lambda^* = \underset{\lambda\in\Lambda}{\arg\min} \ \text{Var}_{g_\lambda}(\hat P(\lambda))=\underset{\lambda\in\Lambda}{\arg\min} \ \text{Var}_{g_\lambda}\left(\II_{\{\varphi(X)\geq 0\}}L_\lambda(X)\right) \]
	with $\hat P(\lambda) = (1/N) \sum_{i=1}^N \II_{\{\varphi(X_i)\geq 0\}}L_\lambda(X_i)$ the IS estimator and $L_\lambda = f / g_\lambda$. The variance minimization problem is not convex with no analytical solution. An alternative and popular choice, which we discuss next, is the cross-entropy method which estimates the parameter~$\lambda$ that minimizes the Kullback-Leibler divergence between $g^*$ and $g_\lambda$.
	
	\subsection{The cross-entropy method (CE)}\label{sub:CE}
	The Kullback-Leibler (KL) divergence between two densities $f$ and $h$ with $f$ absolutely continuous with respect to $h$ is defined by: 
	\[ D(f,h)=\EE_{f}\left[\log \left( \frac{f(X)}{h(X)} \right) \right] = \int f(x) \log \left( \frac{f(x)}{h(x)} \right) \textrm{d} x. \]
	The KL divergence measures a ``distance'' between the two densities. It is always non-negative, and it is zero only when $f = h$ almost everywhere. In order to find an efficient auxiliary density, we can minimize the KL divergence between the optimal $g^*$ and the given parametric auxiliary family and choose $\lambda^*$ given by
	\[ \lambda^*=\underset{\lambda\in\Lambda}{\arg\min} \ D(g^*,g_\lambda). \]
	Since $D(g^*, g_\lambda) = \EE_{g^*}\left[\log(g^*(X))\right] - \EE_{g^*}\left[\log(g_\lambda(X))\right]$ and the first term in the right-hand side is independent of $\lambda$, this problem reduces to maximizing the cross-entropy $\EE_{g^*}\left[\log(g_\lambda(X))\right]$. Using finally that $g^* = f \II_{\{\varphi \geq 0\}}/P$ with $P$ independent of $\lambda$, we get
	\begin{equation} \label{eq:CE-problem}
		\lambda^*=\underset{\lambda\in\Lambda}{\arg\max} \ \EE_f\left[\log\left(g_\lambda(X)\right)\II_{\{\varphi(X)\geq 0\}}\right].
	\end{equation}
	This maximization problem is called the cross-entropy problem. Contrary to the variance minimization problem, the cross-entropy problem is generally concave and differentiable~\cite{rubinstein_cross-entropy_2011, rubinstein_simulation_2017}, so the solution can be found by setting the gradient with respect to $\lambda$ equal to zero. Another advantage is the existence of analytical solutions of the cross-entropy problem for exponential families.
	
	From now on, we suppose that the initial density $f$ is the standard normal density $g_{0, I_n}$ and that the auxiliary density is a Gaussian distribution $\{g_{m, \Sigma}: m \in \RR^n, \Sigma \in \mathcal{M}^+_n\}$, \correction{which belongs to the exponential family}. Here and in the sequel, $g_{m, \Sigma}$ denotes the Gaussian density with mean $m \in \RR^n$ and covariance matrix $\Sigma \in \mathcal{M}^+_n$ with $\mathcal{M}^+_n \subset \RR^{n \times n}$ the set of symmetric, positive-definite matrices, and $I_n$ is the identity matrix in dimension $n$. It is convenient to work in the standard normal space, because it enables easier implementation and computation. By transforming the random inputs X with isoprobabilistic transformations~\cite{liu1986multivariate,hohenbichler1981non}, we can in principle always come back to this case.
	
	In this case, we have $\lambda=(m,\Sigma)\in\RR^{n}\times \mathcal{M}^+_n$ and the cross-entropy maximization problem~\eqref{eq:CE-problem} can be solved explicitly and gives
	\begin{equation}\label{eq:mstar}
	 m^*= \EE_f(X \mid \varphi(X)\geq 0) \ \text{ and } \ \Sigma^*= \EE_f \left( (X-m^*)(X-m^*)^\top \mid \varphi(X)\geq 0 \right),
	\end{equation}
	see for instance~\cite{rubinstein_cross-entropy_2011, geyer_cross_2019}. Since $m^*$ and $\Sigma^*$ depend on the unknown distribution conditional on failure $g^*$, we cannot compute them explicitly but, provided we have a relevant auxiliary density $h$, we can estimate them by IS through
	\begin{equation} \label{eq:hat-m}
		\hat m = \sum_{i=1}^N w_i X_i \ \text{ and } \ \hat \Sigma = \sum_{i=1}^N w_i (X_i-\hat{m})(X_i-\hat{m})^\top
	\end{equation}
	\correction{where} $w_i = w'_i / \sum_j w'_j$ and $w'_i = \II_{\{\varphi(X_i)\geq 0\}} f(X_i) / h(X_i)$ are the unnormalized importance weights, and the $X_i$ are i.i.d.\ samples of density $h$.
	
	To find a good auxiliary density $h$, we can consider sequential simulation techniques to define a sequence of parameters $(m_0, \Sigma_0), (m_1, \Sigma_1), \dots$ Provided $(m_t,\Sigma_t)$ is given, the idea is to generate a sample according to $g_{m_t,\Sigma_t}$ and estimate $(m_{t+1}, \Sigma_{t+1})$ from the empirical $\rho$-quantile ($\rho\in (0,1)$) of $\varphi(X_1),\ldots,\varphi(X_N)$. This is a classical algorithm often referred to as the \correction{multilevel} cross-entropy algorithm (CE). For completeness, a detailed description of CE is provided in \ref{appendix}, Algorithm~\ref{algo:CE} (see also~\cite{rubinstein_cross-entropy_2011}). Without giving all the details recalled in the appendix, it is important to have in mind the following overview of one iteration of CE's main step:
	\begin{enumerate}
		\item Draw a sample $X_1, \ldots, X_N$ from $g_{m_t, \Sigma_t}$ with $(m_t, \Sigma_t)$ the current estimate of $(m^*, \Sigma^*)$;
		\item Compute $\gamma_t$ the empirical $\rho$-quantile of the sample $\varphi(X_1),\ldots,\varphi(X_N)$, and get a new estimation $(m_{t+1}, \Sigma_{t+1})$ of the conditional mean and variance by
		\begin{align}\label{eq:mt+1}
		    m_{t+1} = \sum_{i=1}^N w_i X_i \ \text{ and } \ \Sigma_{t+1} = \sum_{i=1}^N w_i (X_i - m_{t+1})(X_i - m_{t+1})^\top 
		\end{align} 
		\correction{where} $w_i = w'_i / \sum_j w'_j$ and $w'_i = \II_{\{\varphi(X_i)\geq \gamma_t\}} (f/g_{m_t, \Sigma_t})(X_i)$.
	\end{enumerate}
	
	\correction{\begin{rem}[Smooth updating of the parameters]\label{rem:smooth}
	Instead of updating $m_{t+1}$ and $\Sigma_{t+1}$ from \eqref{eq:mt+1}, it is sometimes more efficient to perform a smoothed updating rule, as follows :
	    \[ \widetilde{m}_{t+1} = \alpha m_{t+1}+(1-\alpha)\widetilde{m}_{t}  \ \text{ and } \ \widetilde{\Sigma}_{t+1} = \alpha \Sigma_{t+1}+(1-\alpha)\widetilde{\Sigma}_{t} \]
	where $m_{t+1}$ and $\Sigma_{t+1}$ are estimated in \eqref{eq:mt+1}, $\widetilde{m}_t$ and $\widetilde{\Sigma}_t$ are the smoothed parameters computed at the previous step "t", and $\alpha\in [0,1]$ is the smoothing parameter.
	This procedure can avoid a too fast convergence to degenerate solutions, especially in discrete optimization problems (see \cite{bourinet2018reliability, rubinstein_cross-entropy_2011, costa_convergence_2007} for instance). The smoothing parameter $\alpha$ can be fixed beforehand (static smoothing) or updated at each iteration (dynamic smoothing).
	In the numerical analysis of Section \ref{SectionRes}, we do not perform the smoothing procedure (or equivalently, we keep $\alpha=1$) since it does not improve significantly the estimation results. 
	\end{rem}}

	\subsection{Improved cross-entropy (iCE) method}\label{sub:iCE}
	
	 One of the limitations of CE lies in the estimations of equation~\eqref{eq:hat-m} of the conditional distribution at each step of the algorithm. This estimation only uses a fraction $\rho$ of the sample (recall that $\rho \in (0,1)$ is a parameter of CE, see Section~\ref{sub:CE}), and the idea of iCE is to replace the indicator function $\II_{\{\varphi(X) \geq 0\}}$ by a smooth approximation \correction{as suggested in \cite{papaioannou_improved_2019, katafygiotis2007estimation}, for instance.} \correction{The method proposed in \cite{papaioannou_improved_2019} considers the approximation  $\II_{\{\varphi(x) \geq 0\}} \approx F(\varphi(x) / \sigma)$ with $F$ a cumulative distribution function, typically of the standard normal random variable, so that for $\varphi(x) \neq 0$ we have $F(\varphi(x) / \sigma) \to \II_{\{\varphi(x) \geq 0\}}$ as $\sigma \to 0$. We present this approach with Gaussian densities in Algorithm~\ref{algo:iCE} of ~\ref{appendix}.}
	 
	 \subsection{Comment on the choice of auxiliary family} \label{sub:family}
	 
	 As mentioned above, we restrict our study to the Gaussian auxiliary family $\{ g_{m, \Sigma}: m \in \RR^n, \Sigma \in \mathcal{M}^+_n \}$. This choice is well suited for unimodal problems, which is the typical case that we have in mind, and in such cases CE is known to work well for small dimensions. Our goal here is to improve the efficiency of CE when the dimension increases. In particular, the heuristic arguments justifying CE-$m^*$ presented in Section~\ref{sub:justification} work in this case and not necessarily in more general settings; moreover, we have only considered such test cases for the numerical applications. In case the failure domain is not unimodal, the auxiliary family needs to be updated, see for instance~\cite{papaioannou_improved_2019, wang_cross-entropy-based_2016, geyer_cross_2019}.

	 \section{New algorithms CE-$m^*$ and iCE-$m^*$ with projection} \label{sec:CE-m}
	 
	 \subsection{Inefficiency of CE in high dimension} \label{sub:Ex}
	 
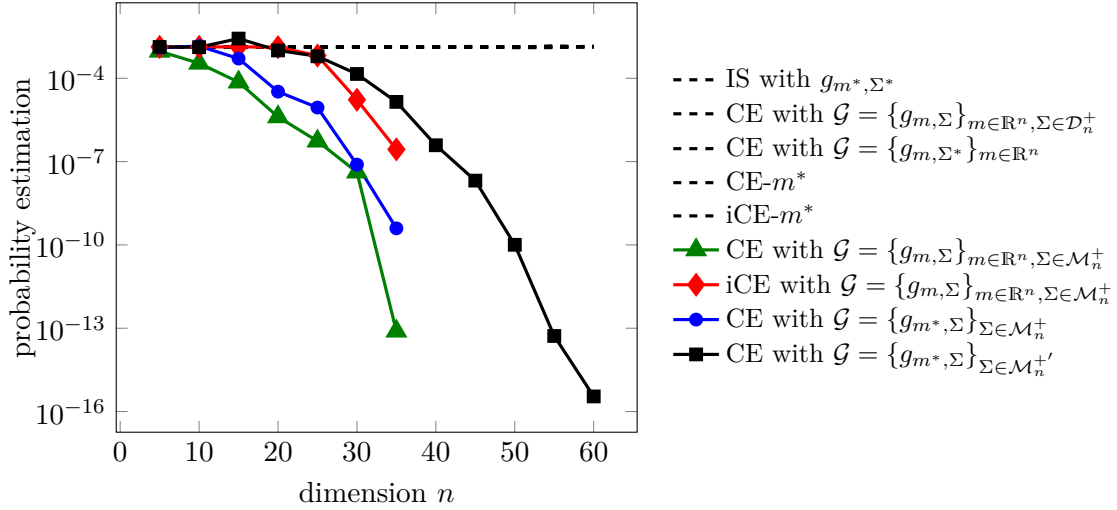
\begin{figure}[t]
			\centering
			\begin{tikzpicture}
				\pgfplotsset{work well/.style={mark=none, dashed, very thick},
					work not/.style={mark=none, solid, very thick}}
			    \begin{axis}[
					xlabel={dimension $n$},
					ylabel={probability estimation},
					legend style={nodes={right}, at={(1.05,.5)}, anchor=west, draw=none, font=\small},
					ymode=log
					]
				    \addplot [work well] table {data/meanISopt2.txt};
				    \addlegendentry{IS with $g_{m^*, \Sigma^*}$}
					\addplot [work well] table {data/meanCEdiago2.txt};
				    \addlegendentry{CE with $\mathcal{G} = \{g_{m, \Sigma}\}_{m \in \RR^n, \Sigma \in \mathcal{D}^+_n}$}
				    \addplot [work well] table {data/meanCE_fixVopt2.txt};
				    \addlegendentry{CE with $\mathcal{G} = \{g_{m, \Sigma^*}\}_{m \in \RR^n}$}
				    \addplot [work well] table {data/meanCEmstar2.txt};
				    \addlegendentry{CE-$m^*$}
				    \addplot [work well] table {data/meaniCEmstar2.txt};
				    \addlegendentry{iCE-$m^*$}
					
				    \addplot [work not, green!50!black, mark=triangle*, mark options={scale=1.8, fill=green!50!black}] table {data/meanCE2.txt};
				    \addlegendentry{CE with $\mathcal{G} = \{g_{m, \Sigma}\}_{m \in \RR^n, \Sigma \in \mathcal{M}^+_n}$}
					\addplot [work not, red, mark=diamond*, mark options={scale=1.8, fill=red}] table {data/meaniCE2.txt};
				    \addlegendentry{iCE with $\mathcal{G} = \{g_{m, \Sigma}\}_{m \in \RR^n, \Sigma \in \mathcal{M}^+_n}$}
				    \addplot [work not, blue, mark=*, mark options={fill=blue}] table {data/meanCE_fixmopt2.txt}; 
				    \addlegendentry{CE with $\mathcal{G} = \{g_{m^*, \Sigma}\}_{\Sigma \in \mathcal{M}^+_n}$}
					\addplot [work not, black, mark=square*, mark options={fill=black}] table {data/meanCE_mixV2.txt};
				    \addlegendentry{CE with $\mathcal{G} = \{g_{m^*, \Sigma}\}_{\Sigma \in {\mathcal{M}}_n^{+'}}$}
			    \end{axis}
			\end{tikzpicture}
			\caption{Evolution of the estimation of the probability $P = \PP_f(\varphi(X) > 0)$ as the dimension increases for $\varphi$ linear given by~\eqref{fct_som} and for different algorithms. Results are obtained with parameters set as in Section~\ref{sub:sum}. In dashed are three algorithms which provide accurate and indistinguishable results: IS with the optimal density $g_{m^*, \Sigma^*}$, CE with auxiliary family $\mathcal{G} = \{g_{m, \Sigma^*}\}_{m \in \RR^n}$ where one uses the optimal variance and estimates the mean, and CE with auxiliary family $\mathcal{G} = \{g_{m, \Sigma}\}_{m \in \RR^n, \Sigma \in \mathcal{D}^+_n}$ where one estimates the mean and only the diagonal of the covariance matrix. In green with triangle is CE with $\{g_{m, \Sigma}\}_{m \in \RR^n, \Sigma \in \mathcal{M}^+_n}$ (both the mean and the full covariance matrix are estimated). In red with diamonds is iCE with the same auxiliary family. In blue with circles is CE with $\{g_{m^*, \Sigma}\}_{\Sigma \in \mathcal{M}^+_n}$ (one uses the optimal mean $m^*$ and estimates the full covariance matrix). In black with squares is CE with $\{g_{m^*, \Sigma}\}_{\Sigma \in \mathcal{M}^{+'}_n}$ (one uses the optimal mean $m^*$ and estimates $3/4$-th of the covariance matrix).}
			\label{fig:inefficiency}
		\end{figure}

	 The inefficiency of CE in high dimension is well documented: the general behavior, even outside a reliability context, is that the estimation of the covariance matrix collapses very suddenly to $0$ \cite{Bengtsson08:0, rubinstein_how_2009, el-laham_robust_2018}.  
	 To illustrate this, let us consider the simple yet illustrative example where $\varphi$ is the following affine function:
	\begin{align}\label{fct_som}
 		\varphi: x=(x_1,\ldots,x_n)\in\RR^n \mapsto \underset{j=1}{\overset{n}{\sum}} x_j - 3 \sqrt{n}
	\end{align}
	such that $P=\PP_f(\varphi(X)\geq 0) = \PP (Z \geq \beta) \simeq 1.35\cdot10^{-3}$ independently of~$n$, with $Z$ following the standard normal distribution on $\RR$. The evolution of the mean probability estimation by CE and iCE is represented in Figure~\ref{fig:inefficiency} \correction{(green triangles for CE, red diamonds for iCE)}.
 		As the dimension increases, the estimation becomes less accurate for both methods. For CE, the error is significant even for dimensions as low as $n \approx 15$, while iCE delivers accurate estimations only up to dimensions $n \approx 25$, after what its accuracy dramatically drops. 
 		Thus even for moderately high dimensions ($n \approx 30$), both methods fail to provide accurate estimations.
		
 		The degradation is due to errors in the parameters estimation. Indeed, both methods aim at finding a good auxiliary density, close to $g^*$, in the family $\{g_{m, \Sigma}\}_{m \in \RR^n, \Sigma \in \mathcal{M}^+_n}$: to do so, they use sequential simulation techniques to get estimates $\hat m$ and $\hat \Sigma$ of the optimal parameters $m^*$ and $\Sigma^*$. As $m^* \in \RR^n$ and $\Sigma^* \in \RR^{n \times n}$ is symmetric, this actually involves the estimation of $n + \frac{1}{2} n(n+1) = \frac{1}{2} n(n+3)$ parameters, a number which grows quickly (quadratically) as the dimension increases. Thus, although the error in each dimension may be low, their global effect may actually be quite significant in the final estimation. This is supported by the numerical results presented in Figure~\ref{fig:inefficiency}. First, we see indeed that IS with the optimal auxiliary density $g_{m^*, \Sigma^*}$ works very well, so final estimation errors indeed come from considering approximations $\hat m$ and $\hat \Sigma$ of $m^*$ and $\Sigma^*$. The four other curves related to CE are somehow in-between the case $g_{m^*, \Sigma^*}$ where $0$ parameter has to be estimated, and the case $\{g_{m,\Sigma}\}_{m \in \RR^n, \Sigma \in \mathcal{M}^+_n}$ where $n + n(n+1)/2$ have to be estimated. In the case $\{g_{m,\Sigma^*}\}_{m \in \RR^n}$ and $\{g_{m,\Sigma}\}_{m \in \RR^n, \Sigma \in \mathcal{D}^+_n}$ one only has to estimate $n$ and $2n$ parameters, respectively ($\mathcal{D}^+_n$ is the set of $n$-dimensional matrices with positive entries on the diagonal and $0$'s elsewhere), and CE with these auxiliary families behave very well. In the two other cases the number of parameters to be estimated grows quadratically in $n$ where we have estimated $P$ with four different methods: among them $\{g_{m^*,\Sigma}\}_{\Sigma \in \mathcal{M}^{+'}_n}$ with $3 n (n+1) / 8$ parameters to estimate behaves better than $\{g_{m^*,\Sigma}\}_{\Sigma \in \mathcal{M}^+_n}$ with $n (n+1) / 2$ parameters to estimate ($\mathcal{M}^{+'}_n$ is the subset of matrices in $\mathcal{M}^+_n$ equal to the identity except for the north-west corner of size $3n/4$). In other words, Figure~\ref{fig:inefficiency} suggests that the efficiency of a given auxiliary family depends on the number of parameters to be estimated: the efficiency improves as the number of parameters to be estimated decreases and a linear number of parameters to estimate behaves very well. This suggests the idea of the present paper, namely lowering the number of estimations needed for the covariance matrix by only estimating the variance in one direction.

		\subsection{Presentation of CE-$m^*$ and iCE-$m^*$} \label{sub:CE-m}
		
		From Figure~\ref{fig:inefficiency} one could be tempted to use the auxiliary family $\{g_{m, \Sigma}\}_{m \in \RR^n, \Sigma \in \mathcal{D}^+_n}$ where one only estimates diagonal entries: however, even if it behaves very well on the particular example of the sum~\eqref{fct_som}, \correction{until dimension 60,} this actually corresponds to a particular choice of auxiliary densities proscribing correlations between input which may in some cases be undesirable. \correction{Moreover, this choice involves more parameter estimation in non influential directions than the proposed method, which can result in inaccurate probability estimation or even the non-convergence of the algorithm in high dimension (see results in Section \ref{SectionRes}). Despite these limitations, we have studied this algorithm numerically as it fits within the framework of the present paper as it can be seen as a dimension-reduction algorithm, see Remark~\ref{rk:diag} below.}
		
		The proposition of this paper is to estimate the variance in the direction of the current estimation of~$m^*$: similarly as the choice $\{g_{m, \Sigma}\}_{m \in \RR^n, \Sigma \in \mathcal{D}^+_n}$ this presents the advantage of only necessitating a linear number of estimations (actually, $n+1$) but in contrast to $\{g_{m, \Sigma}\}_{m \in \RR^n, \Sigma \in \mathcal{D}^+_n}$ this is a more generic choice that does not make any a priori assumption on the problem. 
		 \correction{In fact, an advantage of estimating the variance only in the direction of the updated mean vector is that we do not cancel all the correlation coefficients contrary to the diagonal form of the covariance matrix. The optimal auxiliary IS distribution has dependent components and thus a non diagonal covariance matrix is adapted. Thus, in low dimensions, the lack of flexibility of the proposed model compared to the diagonal covariance matrix is compensated by a better model assumption. The diagonal form assumption constrains the update of the variance in all the directions given by the canonical basis. Among the $n$ directions, very few canonical directions have a variance different from $1$ in high dimension and thus do really need to be updated. On the contrary, in the rare event context, the direction given by the sample mean vector is relevant since the variance has to decrease in this direction even if it is not always the direction that offers the best variance reduction. Having only $1$ covariance matrix parameter to update whatever the dimension $n$ becomes clearly an advantage in high dimensions and explains why CE with diagonal covariance matrix suffers more from the curse of dimensionality than CE-$m^*$, iCE-$m^*$.}
		The precise algorithm is detailed in Algorithm~\ref{algo:CE-m} and its improved (in the sense of iCE) version in Algorithm~\ref{algo:iCE-m}. Its main step can be summarized as follows:
		\begin{enumerate}
			\item Draw a sample $X_1, \ldots, X_N$ from $g_{m_t, \Sigma_t}$ with $(m_t, \Sigma_t)$ the current estimate of $(m^*, \Sigma^*)$;
			\item Compute the empirical $\rho$-quantile $\gamma_t$ of the sample $\varphi(X_1),\ldots,\varphi(X_N)$ and get a new estimation $m_{t+1}$ of the conditional mean $m^*$ by
			\[ m_{t+1} = \sum_{i=1}^N w_i X_i \ \text{ with } \ w_i = \frac{w'_i}{\sum_j w'_j} \ \text{ and } \ w'_i = \II_{\{\varphi(X_i)\geq \gamma_t\}} \frac{f(X_i)}{g_{m_t, \Sigma_t}(X_i)}; \]
			\item Define $Y_i = R^\top X_i$ where $R^\top = m^\top_{t+1} / \lVert m_{t+1} \rVert$: the $Y_i$'s are the projections of the $X_i$'s on $\text{span}(m_{t+1})$, by linearity their weighted mean is equal to $R^\top m_{t+1} = \lVert m_{t+1} \rVert$;
			\item Compute the variance $\hat v$ in the direction of $m_{t+1}$ through
			\[ \hat{v} = \sum_{i=1}^N w_i (Y_i-\lVert m_{t+1}\rVert)^2; \]
			\item Get a new estimation $\Sigma_{t+1}$ of $\Sigma^*$ by considering a variance of $\hat v$ in the direction of $m_{t+1}$ and a unit variance $I_{n-1}$ in its orthogonal: as will be discussed below, this amounts to taking $\Sigma_{t+1} = I_n + (\hat v - 1) R R^\top$.
		\end{enumerate}
		 The only difference between CE (Algorithm~\ref{algo:CE}) and CE-$m^*$ (Algorithm~\ref{algo:CE-m}) is step $9$ of CE-$m^*$: in CE, one estimates the whole covariance matrix through
		\[ \Sigma_{t+1} = \sum_{i=1}^N w_i (X_i-m_{t+1})(X_i-m_{t+1})^\top \]
		whereas in CE-$m^*$, one only estimates a scalar
		\[ \hat{v} = \sum_{i=1}^N w_i (Y_i-\lVert m_{t+1} \rVert)^2 \]
		and then uses $\Sigma_{t+1} = I_n + (\hat v - 1) m_{t+1} m_{t+1}^\top / \lVert m_{t+1} \rVert^2$ (actually, with an additional $\varepsilon I_n$ term, see Remark~\ref{rem:varepsilon} below). The number of parameters to be estimated in CE-$m^*$ thus reduces from $\frac{1}{2} n (n + 3)$ in CE to $n+1$.
		
		\begin{algorithm}[t]
		 \SetAlgoLined
		 \KwData{parameter $\rho\in (0,1)$, sample size $N$ }
		 \KwResult{Estimation $\hat P$ of the failure probability $\PP_f(\varphi(X) \geq 0)$}
		 Initialization: set $t = 0$, $m_t = 0$ and $\Sigma_t = I_n$\;
		 Generate $X_1$, \ldots, $X_N$ independently according to $g_{m_t, \Sigma_t}$ and define $L_t = f/g_{m_t, \Sigma_t}$\;
		 Compute $\varphi_i = \varphi(X_i)$ for $i = 1, \ldots, N$\;
		 Sort them in ascending order : $\varphi_{(1)}\leq \cdots \leq \varphi_{(N)}$, and set $\gamma_t=\varphi_{(\lfloor\correction{(1-\rho)} N\rfloor)}$\;
		 Compute the weights $w_i = w'_i / \sum_j w'_j$ with $w'_i = \II_{\{\varphi_i\geq \gamma_t\}} L_t(X_i)$\;
		 \While{$\gamma_t< 0$}{
		 Compute $m_{t+1} = \sum_{i=1}^N w_i X_i$\;
		 Define $R^\top = m^\top_{t+1}/\lVert m_{t+1}\rVert\in \RR^{1 \times n}$ the projection on $\text{span}(m_{t+1})$\;
		 Compute the projected samples $Y_i = R^\top X_i$ and their conditional variance $\hat{v} = \sum_{i=1}^N w_i (Y_i-\lVert m_{t+1}\rVert)^2$ and set $\Sigma_{t+1}=(\hat{v}-1) R R^\top + (1+\varepsilon) I_n$\;
		 Set $t \longleftarrow t+1$\;
	 	 Repeat Steps 2, 3, 4 and 5 \;
			}
		Return $\displaystyle \hat P=\dfrac{1}{N} \sum_{i=1}^{N} \II_{\{\varphi(X_i)\geq 0\}}L_t(X_i)$.
		 \caption{Proposed algorithm: CE-$m^*$ = CE + projection on span($m^*$).}
		 \label{algo:CE-m}
		\end{algorithm}

		\begin{algorithm}[t]
		 \SetAlgoLined
		 \KwData{parameter $\delta_{\text{target}}$, sample size $N$}
		 \KwResult{Estimation $\hat P$ of the failure probability $\PP_f(\varphi(X) \geq 0)$}
		 Initialization: set $t = 0$, $m_t = 0$, $\Sigma_t=I_n$ and $\sigma_t = \infty$\;
		 Generate $X_1$, \ldots, $X_N$ independently according to $g_{m_t, \Sigma_t}$ and define $L_t = f/g_{m_t, \Sigma_t}$\;
		 Compute $\varphi_i = \varphi(X_i)$ for $i = 1, \ldots, N$\;
		 Compute $\widehat{\text{cv}}$ the empirical coefficient of variation of the $\II_{\{ \varphi_i \geq 0 \}} / F(\varphi_i/\sigma_t)$\;
		 \While{$\widehat{\text{cv}} \geq \delta_{\text{target}}$}{
		 Compute $\displaystyle \sigma_{t+1} = \arg\min ( \hat{\delta}_t(\sigma)- \delta_{\text{target}} )^2 $ where the minimum is taken over $\sigma \in (0,\sigma_t)$ and $\hat{\delta}_t(\sigma)$ is the coefficient of variation of the $F(\varphi_i/\sigma) L_t(X_i)$ with $L_t = f/g_{m_t, \Sigma_t}$\;
		 Compute the weights $w_i = w'_i / \sum_j w'_j$ with $w'_i = F(\varphi_i / \sigma_{t+1}) L_t(X_i)$\;
		 Compute $m_{t+1} = \sum_{i=1}^N w_i X_i$ and set $R = m_{t+1} / \lVert m_{t+1} \rVert$\;
		 Compute the projected samples $Y_i = R^\top X_i$ and their conditional variance $\hat v = \sum_{i=1}^N w_i (Y_i-\lVert m_{t+1}'\rVert)^2$ and set $\Sigma_{t+1}=(\hat v - 1) RR^\top + (1 + \varepsilon) I_n $\;
		 Set $t \longleftarrow t+1$\;
	 	 Repeat Steps 2, 3 and 4\;
			}
		Return $\displaystyle \hat P=\dfrac{1}{N} \sum_{i=1}^{N} \II_{\{\varphi(X_i)\geq 0\}}L_t(X_i)$.
		 \caption{Proposed algorithm: iCE-$m^*$ = iCE + projection on span($m^*$).}
		 \label{algo:iCE-m}
		\end{algorithm}
		
		Having the variance updated in only one direction $d \in \RR^n$ amounts to seeking an auxiliary density of the form $X = \mu + \sigma G d + R_\perp G_\perp$ with $G \sim \NO(0,1)$ and $G_\perp \sim \NO(0, I_{n-1})$ independent standard Gaussian vectors in dimension one and $n-1$, respectively, $\sigma^2$ the variance of $X$ in the direction of $d$ and $R_\perp: \RR^{n-1} \to \RR^n$ that lifts a vector expressed in the basis of $\mathrm{span}(d)^\perp$ (the orthogonal of the vector space spanned by $d$) to $\mathrm{span}(d)^\perp$, which satisfies $d d^\top + R_\perp R_\perp^\top = I_n$. From the expression $X = \mu + \sigma G d + R_\perp G_\perp$, one gets that $X$ follows an $n$-dimensional Gaussian distribution with mean $\mu$ and covariance
		\[ \text{Var}(X) = \sigma^2 d d^\top + R_\perp R^\top_\perp = I_n + (\sigma^2 - 1) d d^\top \]
		which justifies the update rule for $\Sigma_{t+1}$ in step $9$ of CE-$m^*$.
		
		\begin{rem} \label{rem:varepsilon}
			As can be seen in Algorithms~\ref{algo:CE-m} and~\ref{algo:iCE-m}, we actually update $\Sigma_{t+1}$ through $\Sigma_{t+1} = (1+\varepsilon) I_n + (\hat v - 1) R R^\top$: the additional $\varepsilon I_n$ term is here to avoid degeneracy problems due to numerical rounding errors. This numerical trick is also used in the implementation of CE and iCE available on the authors' website \cite{CE_iCE_online}. 
			In all our simulations the value of~$\varepsilon$ is set to $\varepsilon = 10^{-6}$.
		\end{rem}
		
		\correction{\begin{rem}\label{rem:smoothCEm*}
		    As for the CE and iCE algorithms, it is possible to use a smoothed updating procedure (see remark~\ref{rem:smooth}) in CE-$m^*$ and iCE-$m^*$. In this paper, the parameters are not updated with the smoothing rule in the simulations proposed in this article, since it does not improve the results significantly.
		\end{rem}}
		
		\correction{\begin{rem} \label{rk:diag}
			The auxiliary family $\{g_{m, \Sigma}\}_{m \in \RR^n, \Sigma \in \mathcal{D}^+_n}$ amounts to only evaluating variance terms in the $n$ canonical directions of $\RR^n$. As such, it is of the same nature than the algorithms CE-$m^*$ and iCE-$m^*$ that we propose here, i.e., it seeks to improve results by projecting. However, the details differ significantly: although we propose here to project on one adaptive direction, using $\{g_{m, \Sigma}\}_{m \in \RR^n, \Sigma \in \mathcal{D}^+_n}$ amounts to projecting on $n$ static directions. As will be seen below, CE-$m^*$ and iCE-$m^*$ behave better than $\{g_{m, \Sigma}\}_{m \in \RR^n, \Sigma \in \mathcal{D}^+_n}$, which suggests that it is better to have a smaller number of carefully chosen directions, rather than using many directions but which are not adapted to the samples.
		\end{rem}}

\subsection{Justification}\label{sub:justification}
		
		The idea of CE-$m^*$, and other projection algorithms such as~\cite{uribe_cross-entropy-based_2020}, is to only estimate variance parameters in relevant directions. In~\cite{uribe_cross-entropy-based_2020} the choice of the directions is suggested by a theoretical result which provides an upper bound on the KL divergence. However, the computation of these directions requires to compute the gradient of the 
		\correction{limit-state} function which presents two drawbacks: 1) the 
		\correction{limit-state} function needs to be differentiable and 2) computing its gradient is often expensive (and sometimes even out of reach). In comparison, the advantage of CE-$m^*$ is twofold: 1) the gradient need not be computed: this can save a significant amount of simulation budget, and makes the method applicable even if the 
		\correction{performance} function is not differentiable (which is for instance the case for the financial application considered in Section~\ref{sub:portfolio}); and 2) it is extremely simple, in particular there is no additional model evaluation compared to CE. Last but not least, in high dimension where CE behaves poorly, it can significantly improve the performance.
		
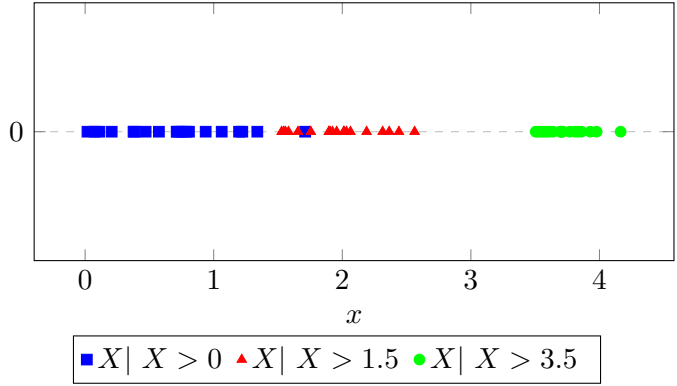
\begin{figure}
\begin{center}
\begin{tikzpicture}
\pgfplotsset{.style={mark=none, solid, very thick}}
\begin{axis}[
    title={},
    height=5cm,
    width=10cm,
    xlabel={$x$},
    ylabel={},
    ytick={0},
    legend style={legend cell align=right, at={(0.5,-.1)}, anchor=north, text=black},
    legend columns=3,
    legend to name=named,
    ymajorgrids=true,
    grid style=dashed,
]

\addplot[only marks,
    color=blue,
    mark=square*,
    ] 
    table{data/Gauss_pts_S0.txt};
    \addplot[only marks,
    color=red,
    mark=triangle*,
    ] 
    table{data/Gauss_pts_S1.txt};
    \addplot[only marks,
    color=green,
    mark=*,
    ]
    table{data/Gauss_pts_S3.txt};
 
  \addlegendentry{ $X | \ X>0$ \ }
	\addlegendentry{ $X | \ X>1.5$ \ }
    \addlegendentry{ $X | \ X>3.5$ \ }
\end{axis}
\end{tikzpicture}

\ref{named}
\end{center}
\caption{\correction{$20$ standard normal samples  conditionally to be greater than three different thresholds in dimension 1.}}
\label{fig:NormSamples}
\end{figure}

		 \correction{At a high level, a good direction in which to project the random variable $X|\varphi(X) \geq 0$ (and thus estimate its variance) is a direction in which its variance is significantly different from the variance of the initial density of $X$, i.e $1$ : indeed, this means that the conditioning by the limit state function plays a role on the resulting variance, in contrast to what happens in directions where the variance remains close to a unit value.} We notice in the simulations that it is indeed the case for the direction $m^*$, i.e., in all the test cases that we considered, the variance in the direction of $m^*$ is systematically smaller than one. The reason for that lies in the very light tail of the Gaussian distribution which makes the variance of the overshoot very small. More precisely, if $X$ is normally distributed, then conditionally on $X > S$ we have for large $S$ the asymptotic expansion $X \approx S + E/S$ with $E$ an exponential random variable. More formally, \correction{by considering Laplace transforms, it is easy to establish that} $S (X - S)$ conditionally on $X > S$ converges in distribution to an exponential random variable from which we can show that the variance of $X$ conditional on $X > S$ goes to $0$ as $S \to \infty$. \correction{We illustrate this situation in Figure \ref{fig:NormSamples}, where we simulate with acceptance-rejection $20$ samples of a standard normal variable $X$ knowing that $X>S$ for an increasing sequence of thresholds $S$. The variance of the resulting samples tends to $0$ as $S$ increases.} 
		
		This behavior is relevant for our purposes because by construction, $m^*$ is the direction to which the CE ``pushes'' the sample, and so heuristically this direction points toward the failure domain. Because of the light tail property of the normal distribution and because the updated variance is only computed on the ``overshoots'', i.e., the points exceeding the border of the failure domain, it is natural that the variance in the direction of $m^*$ decreases. Note that this heuristic argument relies on the unimodal nature of the failure domain (see Section~\ref{sub:family}), since otherwise it can easily be put in default (consider for instance the one-dimensional case $\varphi(x) = \lvert x \rvert - 10$ where the conditional variance actually increases). Note also that this unimodal underlying assumption also intuitively guarantees that the conditional mean $m^*$ is non-zero, thereby leading to a proper definition of the projection on its span.
		
		Consistent with our choice of auxiliary family, it is therefore natural that the variance decreases in the direction of $m^*$, which makes it a good direction to estimate the variance.
		This does not mean that it is \correction{always} the best direction, and we provide indeed an example in Section~\ref{sub:limit} where, for instance, $m^* = \lVert m^* \rVert e_1$ 
		while the best direction (the one in which the variance evolves the most) is $e_2$. However, even in this case, CE-$m^*$ is more efficient than CE in high dimension \correction{(updating the full covariance matrix or only the diagonal)}, see Section~\ref{sub:limit} for more details.

	\section{Numerical Results}\label{SectionRes}
	
	 We compare the two proposed algorithms CE-$m^*$ and iCE-$m^*$ to CE and iCE, \correction{with full covariance matrix and diagonal covariance}, on four examples. \correction{The cross-entropy algorithms updating only the diagonal of the covariance are named CEd and iCEd, while the ones updating the full covariance are still called CE and iCE.} The first example is the linear model already considered above, which illustrates the performance of our algorithms on a simple case where CE has already been shown to fail. The second example considers a modified Ackley function \cite{Ackley_online}, 
	 which is a classical test case in optimization. The third example considers an example from finance (see  \cite{chan_improved_2012,bassamboo_portfolio_2008}), with a 
	 \correction{limit-state} function which is not differentiable. Finally, the last example considers a simple parabolic function that illustrates potential limits of our methods, and thus also ways of improvement.
	 
	  In all simulations we use $\rho=0.1$ for CE and $\delta=1.5$ for iCE, which are classical choices recommended by the authors in \cite{de_boer_tutorial_2005} and \cite{papaioannou_improved_2019}, $\rho=0.1$ for CE-$m^*$ \correction{and CEd}, and $\delta=3$ for iCE-$m^*$ \correction{and iCEd}. For our proposed algorithms, the parameters were selected empirically as the parameters giving the most accurate results. \correction{As mentioned in Remarks \ref{rem:smooth} and \ref{rem:smoothCEm*}, all the algorithms are performed without smoothing.} The sample size may vary from one algorithm to another because the number of iterations may be different, and it is chosen so that the total average simulation budget of the different algorithms is the same \correction{(around 8,000)}. Each algorithm is repeated $100$ times, which produces a sample $\hat P_1, \ldots, \hat P_{100}$ of $100$ independent copies of $\hat P$. In the following tables we report five quantities: the mean estimation, the coefficient of variation, the relative bias, the sample size $N$ and the average simulation budget of these $100$ samples. When $P$ is known they are defined as
	\[ \text{Mean} = \frac{1}{100} \sum_{k=1}^{100} \hat P_k, \ \text{C.o.V.} = \frac{\sqrt{\frac{1}{100} \sum_{k=1}^{100} (\hat P_k - P)^2}}{P} \text{ and } \ \text{Relative bias} = \frac{\text{Mean} - P}{P}, \]
	and when $P$ is unknown then it is first-hand estimated by a crude Monte Carlo method with a high simulation budget.
	
	Moreover, we set the maximum number of iterations of the algorithm to $10$: if after $10$ iterations the algorithm did not stop, we report NC (No Convergence) in the tables. As will be seen below, this is systematically the case for CE and iCE for dimensions $100$ and above. Heuristically, if the algorithm does not converge, this means that it did not manage to build an auxiliary density which samples sufficiently often in the failure domain. This may essentially be due to \correction{three} reasons:
	\begin{enumerate}
		\item either because the estimated mean $m_t$ does not bring the sample close enough to the failure domain: this can typically happen in all \correction{six algorithms} 
		because the variance decreases, so that increments $m_{t+1} - m_t$ become small and the algorithm gets stuck;
		\item \correction{or because the weights are collapsing quickly to 0, hence the estimations in \eqref{eq:hat-m} are not even possible because of divisions by 0; this mainly happens in CEd and iCEd for very high dimension in our examples;} 
		\item or $m_t$ gets close to the failure domain, but the variance does not ``fit'' the failure domain: this can typically happen in CE-$m^*$ and iCE-$m^*$ because $m^*$ is not good enough as a direction of projection (see Remark~\ref{rem:parabola} in Section~\ref{sub:limit}).
	\end{enumerate}
	 
	\correction{We have to note that the non-convergence of these algorithms for the two first reasons can sometimes be avoided with a smoothing procedure or by varying the parameters $\rho$ (for the CE-based methods), and $\delta$ (for the iCE-based methods), but the estimation results stay very inaccurate. Indeed, with a smoothing parameter $\alpha$ between $0.5$ and $0.7$ the algorithms can converge in less than 10 iterations, and with a simulation budget around 8000, but the probability estimation is far from the true probability (often less than $10^{-10}$ instead of $\simeq10^{-3}$). In the same way, decreasing $\rho$ (or increasing $\delta$ for iCE) can make the algorithms converge quickly but the probabilities are almost equal to 0, since the estimations are computed with very small samples (e.g. $\rho N$ with $\rho=0.01$ and $N=4000$). Conversely, increasing $\rho$ (or decreasing $\delta$) does not allow to converge and adds some unnecessary iterations.}
	
	
	\subsection{Toy example: sum of independent variables}\label{sub:sum}

	\begin{table}[h] \small
		\small
	\renewcommand{\arraystretch}{1.2}
	\begin{center}
		\begin{tabular}{|c|c|c|c|c|c|c|c|}
			\hline
			& & iCE & iCE-$m^*$ & \correction{iCEd} & CE & CE-$m^*$ & \correction{CEd} \\
			\hline
			\hline
			\multirow{2}{*}{$n=30$} & 
			Mean ($\times 10^{-3})$ & $0.011$ & $1.35$ & $1.36$ & $5.8\cdot 10^{-5}$ & $1.34$ & 1.35\\
			\cline{2-8}
			\multirow{2}{*}{$P=1.35 \cdot 10^{-3}$} & C.o.V. & $99\%$ & $2.6\%$ & $5.1\%$ & $100\%$ & $4.6\%$ & $5.8\%$
			\\
			\cline{2-8}
			& Relative bias & $-99\%$ & $-0.1\%$ & $0.9\%$ & $-100\%$ & $-0.5\%$ & $0.4\%$\\
			\cline{2-8}
			& Sample size & 1,000 & 2,700 & 3,700 & 1,000 & 2,700 & 3,400 \\\cline{2-8}
			& Simu. budget & 8,500 & 8,000 & 8,000 & 8,300 & 8,100 & 8,000 \\
			\hline
			\hline
			\multirow{2}{*}{$n=100$} & 
			Mean ($\times 10^{-3})$ & NC &  $1.35$ & $1.36$ & NC & $1.35$ & $1.37$\\
			\cline{2-8}
			\multirow{2}{*}{$P=1.35 \cdot 10^{-3}$} & C.o.V. & NC & $4.6\%$ & $6.8\%$ & NC & $4.4\%$ & $7.4\%$
			\\
			\cline{2-8}
			& Relative bias & NC & $0.04\%$ & $0.9\%$ & NC & $-0.08\%$ & $1.1\%$ \\
			\cline{2-8}
			& Sample size & NC & 2,900 & 3,700 & NC & 2,700 & 3,600 \\\cline{2-8}
			& Simu. budget & NC & 8,000 & 8,100 & NC  & 8,100 & 8,100 \\
			\hline
			\hline
			\multirow{2}{*}{$n=300$} & 
			Mean ($\times 10^{-3})$ & NC & $1.35$ & $1.36$ & NC & $1.34$ & $1.31$\\
			\cline{2-8}
			\multirow{2}{*}{$P=1.35 \cdot 10^{-3}$} & C.o.V. & NC & $7.3\%$ & $11.2\%$ & NC & $12.6\%$ & $27.6\%$
			\\
			\cline{2-8}
			& Relative bias & NC & $-0.2\%$ & $0.6\%$ & NC & $-0.8\%$ & $-3.2\%$\\
			\cline{2-8}
			& Sample size & NC & 4,000 & 3,700 & NC & 2,700 & 3,700 \\\cline{2-8}
			& Simu. budget & NC & 8,000 & 8,000 & NC & 8,100 & 8,000 \\
			\hline
		\end{tabular}
	\end{center}
	\caption{Simulation results when $\varphi$ is given by the sum of coordinates~\eqref{fct_som}. Recall that NC means no convergence, see beginning of Section~\ref{SectionRes}.}
	\label{Table:sum}
	\end{table}

		First we come back to the linear case~\eqref{fct_som} of Section~\ref{sub:Ex} where $\varphi(x) \approx \sum_{j=1}^n x_j - 3 \sqrt n$ and $P=\PP(\varphi(X) \geq 0) = 1.35\cdot10^{-3}$, for all dimension~$n$. We already saw in~Figure~\ref{fig:inefficiency} that the performance of CE and iCE degrades significantly as the dimension~$n$ increases. Figure~\ref{fig:inefficiency} also presents results for CE-$m^*$, iCE-$m^*$, \correction{CEd and iCEd} (in dotted lines) which are indistinguishable from those obtained by, for instance, CE with the optimal auxiliary density $g_{m^*, \Sigma^*}$. This shows that projecting on $m^*$ stabilizes the estimation up to the dimension $n = 60$ considered in this figure. \correction{However, CEd and iCEd have also similar performance on this test case.}
	
		We report in Table~\ref{Table:sum} quantitative results for dimensions $n = 30, 100$ and $300$. As suggested by Figure~\ref{fig:inefficiency}, even for dimension $n = 30$ both CE and iCE provide inaccurate estimates with almost $-100\%$ relative bias, meaning that almost no sample falls in the failure domain, whereas for the same dimension, the relative bias of 
		\correction{the other algorithms} is less than $1\%$ with coefficients of variation between $2$ and $6\%$. In dimensions $n = 100$ and $300$, CE and iCE fail to converge. In sharp contrast, the performance of CE-$m^*$ and iCE-$m^*$ remains very good, with relative bias less than $1\%$ and coefficient of variation between $4$ and $13\%$. \correction{The C.o.V. and the relative bias of CEd and iCEd are slightly larger than for CE-$m^*$ and iCE-$m^*$ respectively, in dimension $n=100$. For $n=300$, iCEd is again slightly less efficient than iCE-$m^*$ (C.o.V. equal to 11.2$\%$ against $7.3\%$, and similar relative bias), and CEd is clearly less accurate than CE-$m^*$ since the coefficient of variation is twice larger. We suggest that CE-$m^*$ is better than CEd in high dimension because CEd updates more parameters which are not influential on the estimation and involve more estimation errors whereas CE-$m^*$ updates the variance in the optimal direction.} 
		\correction{We note that on this example, iCE-$m^*$ and iCEd perform better than CE-$m^*$ and CEd respectively, with lower or comparable coefficient of variation and lower relative bias. Moreover, iCE-$m^*$ converges faster (with less iterations) which makes its sample size larger in dimension $n = 300$.}

	\subsection{Application in optimization: modified Ackley function}\label{sub:ackley}
	
		\begin{table}[h] \small
		\small
	\renewcommand{\arraystretch}{1.2}
	\begin{center}
		\begin{tabular}{|c|c|c|c|c|c|c|c|}
			\hline
			& & iCE & iCE-$m^*$ & \correction{iCEd} & CE & CE-$m^*$ & \correction{CEd} \\
			\hline
			\hline
			\multirow{2}{*}{$n=30$} & 
			Mean ($\times 10^{-3})$ & $1.23$ & $1.62$ & $1.65$ & $0.92$ & $1.62$ & 1.66\\
			\cline{2-8}
			\multirow{2}{*}{$P=1.64\cdot 10^{-3}$} & C.o.V. & $89\%$ & $7.6\%$ & $9.0\%$ & $75\%$ & $13.2\%$ & $9.9\%$
			\\
			\cline{2-8}
			& Relative bias & $-24.7\%$ & $-1.1\%$ & $0.6\%$ & $-43.9\%$ & $-1.2\%$ & $0.9\%$\\
			\cline{2-8}
			& Sample size & 1,000 & 2,700 & 2,700 & 2,700 & 2,200 & 2,700 \\\cline{2-8}
			& Simu. budget & 8,100 & 8,000 & 8,100 & 8,100 & 8,000 & 8,100 \\
			\hline
			\hline
			\multirow{2}{*}{$n=100$} & 
			Mean ($\times 10^{-3})$ & NC &  $1.17$ & $1.19$ & NC & $1.22$ & $1.20$\\
			\cline{2-8}
			\multirow{2}{*}{$P=1.18 \cdot 10^{-3}$} & C.o.V. & NC & $13.0\%$ & $10.0\%$ & NC & $29.4\%$ & $21.4\%$
			\\
			\cline{2-8}
			& Relative bias & NC & $-1.2\%$ & $0.9\%$ & NC & $3.4\%$ & $1.6\%$ \\
			\cline{2-8}
			& Sample size & NC & 2,700 & 2,700 & NC & 2,200 & 2,700 \\\cline{2-8}
			& Simu. budget & NC & 8,100 & 8,100 & NC  & 8,000 & 8,100 \\
			\hline
			\hline
			\multirow{2}{*}{$n=200$} & 
			Mean ($\times 10^{-3})$ & NC & $1.71$ & $1.69$ & NC & $1.72$ & $1.82$\\
			\cline{2-8}
			\multirow{2}{*}{$P=1.72 \cdot 10^{-3}$} & C.o.V. & NC & $9.1\%$ & $11.6\%$ & NC & $26.9\%$ & $120\%$
			\\
			\cline{2-8}
			& Relative bias & NC & $-0.5\%$ & $-1.5\%$ & NC & $-0.05\%$ & $5.7\%$\\
			\cline{2-8}
			& Sample size & NC & 2,700 & 2,700 & NC & 2,700 & 3,700 \\\cline{2-8}
			& Simu. budget & NC & 8,100 & 8,000 & NC & 8,100 & 8,000 \\
			\hline
		\end{tabular}
	\end{center}
	\caption{Simulation results when $\varphi$ is given by the modified Ackley function~\eqref{eq:modified-Ackley}.}
	\label{Table:ackley}
	\end{table}

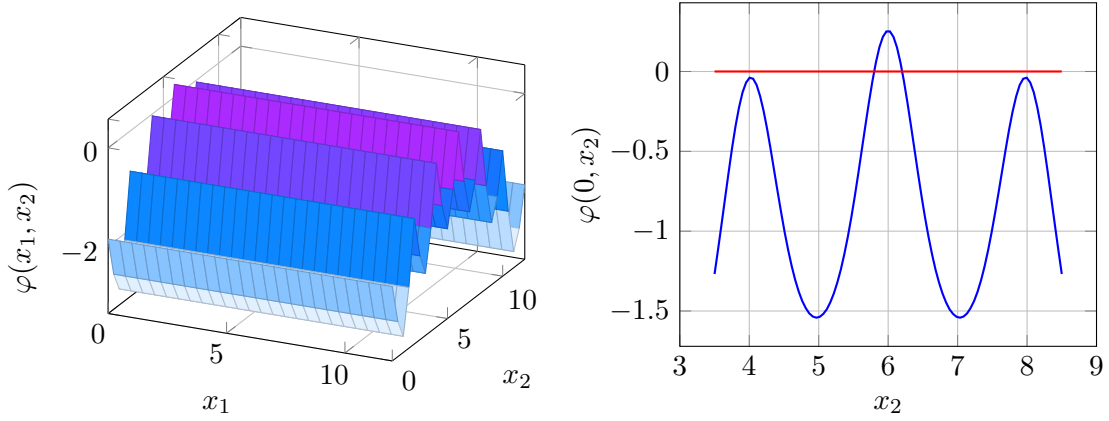
\begin{figure}[h]
\begin{center}
\begin{tikzpicture}
\begin{axis}[
    axis lines= box,
    colormap/cool,
    xlabel={$x_1$},
    ylabel={$x_2$},
    zlabel={$\varphi(x_1,x_2)$},
    grid,
    scale=0.8
]

\addplot3[
    surf,
    domain=0:12,
]
{20*exp(-0.2*sqrt((0*x-3)^2/2+(0.5*y-3)^2/2))+exp(cos(deg(2*pi*(0*x-3)))/2+cos(deg(2*pi*(0.5*y-3)))/2)-15.55};

\end{axis}
\end{tikzpicture}
\hskip 2pt
\begin{tikzpicture}
\begin{axis}[
    xlabel={$x_2$},
    ylabel={$\varphi(0,x_2)$},
    grid,
    samples=100,
    scale=0.8
]

\addplot[domain=3.5:8.5, color=blue, thick
]
{20*exp(-0.2*sqrt((-3)^2/2+(0.5*x-3)^2/2))+exp(cos(deg(2*pi*(-3)))/2+cos(deg(2*pi*(0.5*x-3)))/2)-15.55};
\addplot[domain=3.5:8.5, color=red, thick]{0};
\end{axis}
\end{tikzpicture}
\end{center}
\caption{\correction{2-dimensional modified Ackley function \ref{eq:modified-Ackley} (left) and the projection in the plane $\{x_1=0\}$ (right). The failure threshold is represented by the red line.}}
\label{fig:ackley}
\end{figure}

	The Ackley function \cite{Ackley_online} 
	is a non-convex function widely used to test optimization algorithms that we use here to test the performance of CE-$m^*$ and iCE-$m^*$ against the dimension. Thus we consider
\begin{equation} \label{eq:modified-Ackley}
	\varphi(x) = 20 \exp\left(-0.2\sqrt{\frac{1}{n}\underset{j=1}{\overset{n}{\sum}} (a_jx_j-3)^2}\right) + \exp\left(\frac{1}{n}\underset{j=1}{\overset{n}{\sum}} \cos\left(2\pi (a_jx_j-3)\right)\right) - c_n
\end{equation} 
with $a=(a_1,\ldots,a_n)=\frac{2}{n}(0,1,2,\ldots,n-1)\in\RR^n$ a fixed vector and $c_n$ a \correction{real-valued} constant depending on the dimension that we adjust in order to choose the value of the probability. The standard Ackley function corresponds to $a_i = 1$ for all $i$: for this choice, symmetry considerations imply that $m^* =\lVert m^*\rVert (1,\dots,1)$ is the vector where all coordinates are equal. This is the same as in the previous example, and we introduce the additional parameter $a$ in order to break this symmetry. 
Indeed, we can numerically estimate $m^*=(m_1^*,\ldots,m^*_n)$, for instance by brute force Monte Carlo with a large simulation budget, and we get $m_1^*\approx 0.01 < m^*_2 < \cdots < m^*_n\approx 0.46$.
\correction{In addition, the function does not depend on the first variable $x_1$, since $a_1=0$.} 
\correction{We plot the 2-dimensional modified Ackley function in Figure \ref{fig:ackley}, and its projection in the plane $\{x_1=0\}$. The failure domain corresponds to the points $(x_1,x_2)$ such that $\varphi(x_1,x_2)\geq 0$, i.e. values of $x_2$ such that the blue curve is above the red line on the right graph of Figure \ref{fig:ackley}, since the function does not depend on $x_1$.  }


The results are given in Table~\ref{Table:ackley} and are qualitatively similar to those of the previous example with the sum: for dimension $n = 30$, CE and iCE give poor results (although better than for the sum, the absolute relative bias being between $25$ and $45\%$) while CE-$m^*$, iCE-$m^*$, \correction{CEd, and iCEd} give good estimation with a relative bias of $1\%$ or less and coefficient of variation between~$7$ and $13\%$. As the dimension increases CE and iCE do not converge and the results for CE-$m^*$, iCE-$m^*$, \correction{and iCEd} stay qualitatively similar: the relative bias remains small, between $1$ and $3\%$, and the coefficient of variation between $9$ and $30\%$. \correction{CEd is also very close to CE-$m^*$ in dimension $n=30$ and 100, and even a little more accurate (since the C.o.V. and the relative bias are smaller), however, in dimension $n=200$, CEd has large coefficient of variation ($120\%$) and relative bias ($5.7\%$) contrary to CE-$m^*$. Then, CEd seems to be less efficient than CE-$m^*$ in high dimension because it estimates too many non-influential parameters.} Similarly as for the sum, we note that \correction{the iCE-based methods perform better than the ones based on CE.} 

	\subsection{Application in finance: large portfolio losses}\label{sub:portfolio}
	
	\begin{table}[h] \small
		\small
	\renewcommand{\arraystretch}{1.2}
	\begin{center}
		\begin{tabular}{|c|c|c|c|c|c|c|c|}
			\hline
			& & iCE & iCE-$m^*$ & \correction{iCEd} & CE & CE-$m^*$ & \correction{CEd} \\
			\hline
			\hline
			\multirow{2}{*}{$n=30$} & 
			Mean ($\times 10^{-3})$ & $0.0028$ & $4.29$ & $4.38$ & $2.7$ & $4.24$ & 4.32\\
			\cline{2-8}
			\multirow{2}{*}{$P=4.29\cdot 10^{-3}$} & C.o.V. & $100\%$ & $10.1\%$ & $8.1\%$ & $54\%$ & $7.2\%$ & $9.0\%$
			\\
			\cline{2-8}
			& Relative bias & $-100\%$ & $-0.06\%$ & $2.1\%$ & $-37\%$ & $-1.3\%$ & $0.8\%$\\
			\cline{2-8}
			& Sample size & 1,000 & 2,700 & 3,200 & 2,600 & 2,700 & 2,700 \\\cline{2-8}
			& Simu. budget & 8,400 & 7,900 & 8,000 & 8,000 & 8,100 & 8,100 \\
			\hline
			\hline
			\multirow{2}{*}{$n=100$} & 
			Mean ($\times 10^{-3})$ & NC &  $1.9$ & $1.8$ & NC & $1.8$ & $1.8$\\
			\cline{2-8}
			\multirow{2}{*}{$P=1.8 \cdot 10^{-3}$} & C.o.V. & NC & $8.5\%$ & $7.1\%$ & NC & $8.4\%$ & $14.4\%$
			\\
			\cline{2-8}
			& Relative bias & NC & $3.0\%$ & $2.0\%$ & NC & $1.3\%$ & $-1.8\%$ \\
			\cline{2-8}
			& Sample size & NC & 3,000 & 2,700 & NC & 2,700 & 2,700 \\\cline{2-8}
			& Simu. budget & NC & 8,000 & 8,000 & NC  & 8,100 & 8,100 \\
			\hline
			\hline
			\multirow{2}{*}{$n=250$} & 
			Mean ($\times 10^{-5})$ & NC & $1.01$ & $0.049$ & NC & $0.94$ & NC\\
			\cline{2-8}
			\multirow{2}{*}{$P=1.0 \cdot 10^{-5}$} & C.o.V. & NC & $110\%$ & $98\%$ & NC & $60\%$ & NC
			\\
			\cline{2-8}
			& Relative bias & NC & $1.0\%$ & $-95\%$ & NC & $-5.9\%$ & NC\\
			\cline{2-8}
			& Sample size & NC & 1,600 & 1,500 & NC & 2,100 & NC \\\cline{2-8}
			& Simu. budget & NC & 8,000 & 8,000 & NC & 8,100 & NC \\
			\hline
		\end{tabular}
	\end{center}
	\caption{Simulation results when $\varphi$ is given by the portfolio loss function~\eqref{eq:varphi-loss}.
	}
	\label{Table:portfolio}
	\end{table}

	Next we consider a financial application taken from~\cite{chan_improved_2012, bassamboo_portfolio_2008}. In this example, the failure probability is $P=\PP(L(Z)>\correction{b n})$, \correction{where} $L$ is a loss function defined by
\[ L(z)=\underset{j=1}{\overset{n}{\sum}}\II_{\{z_j\geq 0.5\sqrt{n}\}}. \]
\correction{We choose $b=0.45$ when $n=30$, and $b=0.25$ when $n=100$ and $250$.}
The entries $Z_j$ are dependent real variables defined by 
\[ Z_j = \left( q U+ (1-q^2)^{1/2} \eta_j \right) \mu^{-1/2} \]
where $U\sim\NO(0,1)$, $\eta_j\sim\NO(0,9)$, $j=1,\ldots,n$, $\mu\sim\text{Gamma}(6,6)$ are all independent variables, and $q = 0.25$. \correction{The same realisation of $U$ and $\mu$ is considered for the computation of all the variables $Z_j$.}
%
In our methods, as we work only with normal distributions and our initial density $f$ has to be \correction{standard normal}, we set $\eta_j=3\widetilde{\eta}_j$, for all $j=1,\ldots n$ and $\mu=F_\Gamma^{-1}(F_\NO(\widetilde{\mu}))$ with $\widetilde{\eta}_j$, $\widetilde{\mu}$ independent standard normal variables and $F_\Gamma$, $F_\NO$ the cumulative distribution functions of the $\text{Gamma}(6,6)$ and $\NO(0,1)$ respectively. Hence, if we set $X_1 = U, X_2 = \widetilde \mu$ and $(X_3, \ldots, X_{n+2}) = \widetilde \eta \in \RR^n$ and we define
\begin{align} \label{eq:varphi-loss}
	\varphi(X) = \underset{j=1}{\overset{n}{\sum}} \II_{\{\psi(U, \widetilde \mu, \widetilde \eta_j) \geq 0.5\sqrt{n}\}}-\correction{b n}
\end{align}
with
\[ \psi(U,\widetilde{\mu},\widetilde{\eta}_j) = \left( q U + 3 (1-q^2)^{1/2} \widetilde{\eta}_j \right) \left[ F_\Gamma^{-1} \left( F_\NO(\widetilde{\mu}) \right) \right]^{-1/2} \]
then the sought probability $P = \PP(L(Z) > \correction{b n})$ gets rewritten as $P = \PP(\varphi(X)>0)$ with $X$ a standard Gaussian vector in dimension $n+2$. Then we test the performance of CE-$m^*$ and iCE-$m^*$ against CE, iCE, \correction{CEd and iCEd} for dimension $n=30$, $100$ and $250$. Results are gathered in Table~\ref{Table:portfolio}.

Similarly as in the previous examples, iCE fails in dimension $n = 30$ while CE leads to an inaccurate estimate with a relative bias of $-37\%$ and a coefficient of variation of~$54\%$. For the same dimension and with a similar simulation budget, CE-$m^*$, iCE-$m^*$, \correction{CEd and iCEd} give an estimation with relative bias less than $2\%$ and coefficient of variation less than~$10\%$. In dimension $n = 100$, CE and iCE fail to converge while 
\correction{the four other algorithms} remain accurate with a relative bias less than $3\%$ and a coefficient of variation of $8\%$, \correction{except CEd which has $14.4\%$ of C.o.V. and is slightly less accurate}. Although CE-$m^*$ and iCE-$m^*$ behave similarly in dimension $n = 30$ and $100$, in dimension $n = 250$ CE-$m^*$ behaves better than iCE-$m^*$ with a comparable relative bias ($6\%$ vs.\ $1\%$) but a coefficient of variation which is twice smaller ($60\%$ vs.\ $110\%$). \correction{On the other hand, in dimension $n=250$, we observe that CEd doesn't converge because of the fast weight degeneracy to 0, and iCEd fail to estimate accurately the probability with a relative bias of $-95\%$. }

\subsection{Example where $m^*$ is not the best direction} \label{sub:limit}

\begin{figure}[h]
    \centering
    \includegraphics[width=1\textwidth]{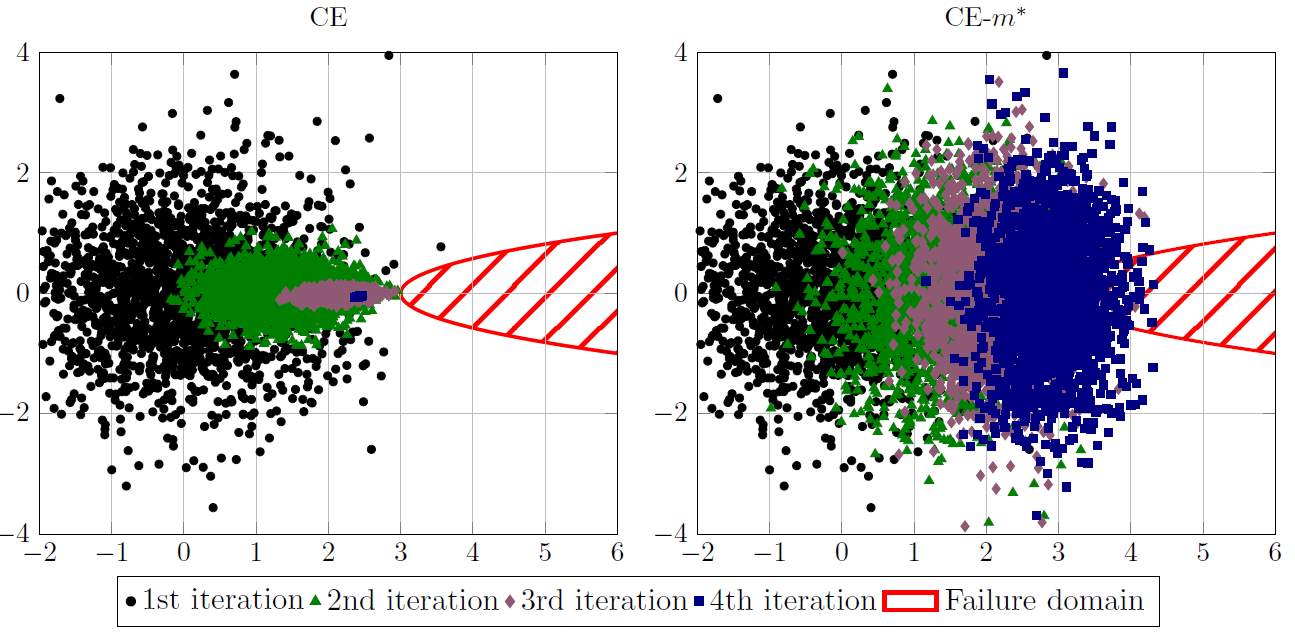}
    \caption{Projection on the $(x_1, x_2)$ plane of the failure domain of the function $\varphi(x) = x_1 -  3x^2_2 - 3$ considered in \eqref{eq:parabola} and the samples of every iteration of CE (left) and CE-$m^*$ (right), with dimension $n=100$. Since $\varphi$ does not depend on other coordinates, the full failure domain is a tube around this two-dimensional shape.}
    \label{fig:parabola}
\end{figure}

	\begin{table}[h] \small
		\small
	\renewcommand{\arraystretch}{1.2}
	\begin{center}
		\begin{tabular}{|c|c|c|c|c|c|c|c|}
			\hline
			& & iCE & iCE-$m^*$ & \correction{iCEd} & CE & CE-$m^*$ & \correction{CEd} \\
			\hline
			\hline
			\multirow{2}{*}{$n=30$} & 
			Mean ($\times 10^{-4})$ & $2\cdot 10^{-6}$ & $2.9$ & $2.9$ & $9\cdot10^{-8}$ & $2.9$ & 2.6\\
			\cline{2-8}
			\multirow{2}{*}{$P=2.9\cdot 10^{-4}$} & C.o.V. & $100\%$ & $11.4\%$ & $5.2\%$ & $100\%$ & $11.2\%$ & $37\%$
			\\
			\cline{2-8}
			& Relative bias & $-100\%$ & $1.0\%$ & $0.2\%$ & $-100\%$ & $0.5\%$ & $-11\%$\\
			\cline{2-8}
			& Sample size & 1,000 & 2,700 & 2,700 & 1,000 & 1,600 & 2,000 \\\cline{2-8}
			& Simu. budget & 8,800 & 8,000 & 8,100 & 8,800 & 7,900 & 8,000 \\
			\hline
			\hline
			\multirow{2}{*}{$n=100$} & 
			Mean ($\times 10^{-4})$ & NC &  $2.9$ & $2.9$ & NC & $3.0$ & $1.4$\\
			\cline{2-8}
			\multirow{2}{*}{$P=2.9 \cdot 10^{-4}$} & C.o.V. & NC & $11.3\%$ & $7.2\%$ & NC & $28.3\%$ & $87\%$
			\\
			\cline{2-8}
			& Relative bias & NC & $1.1\%$ & $-0.3\%$ & NC & $1.3\%$ & $-52\%$ \\
			\cline{2-8}
			& Sample size & NC & 2,700 & 2,600 & NC & 1,900 & 2,000 \\\cline{2-8}
			& Simu. budget & NC & 8,000 & 8,000 & NC  & 8,100 & 8,000 \\
			\hline
			\hline
			\multirow{2}{*}{$n=300$} & 
			Mean ($\times 10^{-4})$ & NC & $3.0$ & NC & NC & $2.9$ & NC \\
			\cline{2-8}
			\multirow{2}{*}{$P=2.9 \cdot 10^{-4}$} & C.o.V. & NC & $29.2\%$ & NC & NC & $88\%$ & NC
			\\
			\cline{2-8}
			& Relative bias & NC & $3.5\%$ & NC & NC & $-1.5\%$ & NC\\
			\cline{2-8}
			& Sample size & NC & 2,300 & NC & NC & 2,400 & NC \\\cline{2-8}
			& Simu. budget & NC & 8,100 & NC & NC & 8,000 & NC \\
			\hline
		\end{tabular}
	\end{center}
	\caption{Simulation results when $\varphi$ is given by the parabola function \eqref{eq:parabola}.
	}
	\label{Table:parabola}
	\end{table}

We conclude this numerical section with an example showing the robustness of CE-$m^*$ and iCE-$m^*$. This is an example where, although $m^*$ is clearly not the best direction in which to estimate the variance, they do still improve on CE and iCE in high dimension. Consider the function
\begin{align}\label{eq:parabola}
    \varphi(x) = x_1 - 3 x^2_2 - 3, \ x \in \RR^n 
\end{align} 
whose failure domain is represented in Figure~\ref{fig:parabola}. By symmetry it is clear that under the optimal auxiliary density $g^*$, which is the distribution conditional on failure, the average~$m^*$ is of the form $m^* = \alpha e_1$ with $\alpha \approx 3.4$ numerically evaluated and the variance $\Sigma^*$ is equal to the identity, except for the variance terms associated to the first two coordinates, which can be numerically estimated to  $\Sigma^*_{11} \approx 0.095$ and $\Sigma^*_{22} \approx 0.044 $. Once the mean is within the failure domain, it is clear that one should decrease the variance in the direction of $e_2$ because the failure domain is quite narrow in this direction, as is reflected by the fact that $\Sigma^*_{22} < \Sigma^*_{11}$. And indeed, one can check that IS with such an auxiliary density behaves better. However, CE-$m^*$ only updates the variance in the (approximate) direction of~$e_1$ and not $e_2$, so the density is not optimal to estimate the probability of failure, as we can see in Figure~\ref{fig:parabola}. Nevertheless, in high dimension CE-$m^*$ converges and is able to estimate the probability whereas CE doesn't because of the fast covariance matrix shrinkage. From a theoretical point of view, we do not fully understand why the covariance matrix collapses for CE and CE-$m^*$. Preliminary investigation suggests that this has to do with errors made in covariance terms, which are larger for CE than CE-$m^*$. The reason why CE-$m^*$ avoids the covariance matrix shrinkage problem constitutes in our view an interesting research question.

We can see it in the simulation results given in Table~\ref{Table:parabola}. Similarly to the previous examples, in dimension $n=30$, CE and iCE already give inaccurate estimations with $100\%$ of coefficient of variation and $-100\%$ of relative bias, whereas CE-$m^*$ and iCE-$m^*$ are very efficient (around $11\%$ of coefficient of variation and between $0.5$ and $1\%$ of relative bias). \correction{iCEd is even more accurate with $5.2\%$ of C.o.V. and $0.2\%$ of relative bias but CEd isn't so efficient since its coefficient of variation is $37\%$ and its relative bias is $-11\%$.} For higher dimension, CE and iCE fail to converge. 
\correction{In dimension $n=100$, iCE-$m^*$ and iCEd are still very accurate with a coefficient of variation between $7$ and $12\%$, whereas CE-$m^*$ and CEd gives $28.3\%$ and $87\%$ of C.o.V. respectively. CEd is particularly inaccurate since its relative bias is $-52\%$, while the three other convergent algorithms have a relative bias lower than $1.3\%$. For dimension $n=300$, iCEd and CEd doesn't converge anymore, because of the weight degeneracy, whereas CE-$m^*$ and iCE-$m^*$ are still able to estimate the probability with a small relative bias ($-1.5\%$ and $3.5\%$ respectively), although the coefficient of variation can be large, especially for CE-$m^*$ ($88\%$ against $29\%$ for iCE-$m^*$).} 

Hence, although $m^*$ is not the optimal direction to estimate the covariance matrix, this example shows that CE-$m^*$ and iCE-$m^*$ are able to efficiently estimate the probability in high dimension while CE, iCE, \correction{CEd and iCEd} fail.

\begin{rem}\label{rem:parabola}
    If we consider a narrower parabola, (for example with $\varphi(x)=x_1-25x_2^2-3$), then CE-$m^*$ is not converging anymore. Indeed, in such cases the final density does not ``fit'' the failure domain and most of the samples fall outside the domain. This happens because $m^*$ is not the best direction in which updating the covariance matrix. So this is a case where CE-$m^*$ does not improve CE performance. \correction{However, CE, CEd, iCE and iCEd do not converge either in high dimension on this example.} Moreover, in high dimension, we did not find any test case (satisfying the unimodal assumption) where CE-$m^*$ degrades CE performance. And as shown in the above numerical examples, most of the times it greatly improves it.
\end{rem}

 \section{Conclusion}\label{Concl}

In this paper, we present a new CE-based importance sampling method using a one-dimensional projection that provides accurate estimation of rare event probabilities in high dimension, provided the target optimal auxiliary density is unimodal. At each step of the algorithm, the idea is to only estimate variance and covariance terms in the direction of the current estimate of the best auxiliary parameter $m^*$: the light tail of the Gaussian distribution guarantees that the variance decreases in this direction which ensures that there will be more information than noise in estimating the variance in this direction. Empirically, our results show that, in high dimension, it is much more efficient to only estimate variance terms in one ``good'' direction than blindly in the whole space as is done in CE, \correction{ or to estimate the variance diagonal terms (as in CEd)}. Although extremely simple, our algorithm turns out to be very efficient, all the more considered that it does not require gradient estimation. Indeed, we find that it cannot degrade CE performance and that in most cases, it actually greatly improves it, typically providing accurate estimates where CE simply failed to converge. \correction{Moreover, in moderately high dimension (say until 100), CE-$m^*$ is as efficient as CEd, and it is often more accurate in higher dimension (more than 200).}

This method opens many interesting research direction which we aim to pursue. First, it would be interesting to get stronger theoretical justification of the accuracy and robustness of our algorithm. It seems in particular that it avoids the well-known shrinkage problem of CE because it estimates less covariance terms, and it would be interesting to confirm this hypothesis. Moreover, the intuition behind the choice of $m^*$ as a direction to project could be deepened to select alternative, possibly better directions. The intuition being that the variance evolves in the direction of $m^*$, this naturally suggests to look at specific eigenvalues of the covariance matrix, which we are currently investigating. Finally, it would be interesting to extend our approach to the multi-modal case by trying to combine it with existing algorithms developed in this case in low dimensions.

\appendix

	\section{} \label{appendix}
	
	In this appendix we simply recall the basic CE and iCE algorithms.

	 \begin{algorithm}[h!]
	 \SetAlgoLined
	 \KwData{parameter $\rho \in (0,1)$, and sample size $N$ }
	 \KwResult{Estimation $\hat P$ of the failure probability $\PP_f(\varphi(X) \geq 0)$}
	 Initialization: set $t = 0$, $m_t = 0$ and $\Sigma_t=I_n$\;
	 Generate $X_1$, \ldots, $X_N$ independently according to $g_{m_t, \Sigma_t}$\;
	 Compute $\varphi_i = \varphi(X_i)$\;
	 Sort the samples in ascending order: $\varphi_{(1)} \leq \dots \leq \varphi_{(N)}$, and set $\gamma_t = \varphi_{(\lfloor \correction{(1-\rho)} N \rfloor)}$\;
	 Compute the weights $w_i = w'_i / \sum_{j=1}^n w'_j$ with $w'_i = \II_{\{\varphi_i \geq \gamma_t\}} L_t(X_i)$\;
	 \While{$\gamma_t < 0$}{
	 Compute $m_{t+1} = \sum_{i=1}^N w_i X_i$ and $\Sigma_{t+1} = \sum_{i=1}^N w_i (X_i-m_{t+1})(X_i-m_{t+1})^\top$\;
	 Set $t \longleftarrow t+1$\;
	 	 Repeat Steps 2, 3, 4 and 5\;
	 	}
	 	Return $\displaystyle \hat P=\dfrac{1}{N} \sum_{i=1}^{N} \II_{\{\varphi(X_i)\geq 0\}}L_t(X_i)$.
	 \caption{CE method \cite{rubinstein_cross-entropy_2011}}
	 \label{algo:CE}
	 \end{algorithm}
	
	\begin{algorithm}[h!]
	 \SetAlgoLined
	 \KwData{parameter $\delta_{\text{target}}$, sample size $N$}
	 \KwResult{Estimation $\hat P$ of the failure probability $\PP_f(\varphi(X) \geq 0)$}
	 Initialization: set $t = 0$, $m_t = 0$, $\Sigma_t=I_n$ and $\sigma_t = \infty$\;
	 Generate $X_1$, \ldots, $X_N$ independently according to $g_{m_t, \Sigma_t}$\;
	 Compute $\varphi_i = \varphi(X_i)$ for $i = 1, \ldots, N$\;
	 Compute $\widehat{\text{cv}}$ the empirical coefficient of variation of the $\II_{\{ \varphi_i \geq 0 \}} / F(\varphi_i/\sigma_t)$\;
	 \While{$\widehat{\text{cv}} \geq \delta_{\text{target}}$}{
	 Compute $\displaystyle \sigma_{t+1} = \arg\min ( \hat{\delta}_t(\sigma)- \delta_{\text{target}} )^2 $ where the minimum is taken over $\sigma \in (0,\sigma_t)$ and $\hat{\delta}_t(\sigma)$ is the coefficient of variation of the $F(\varphi_i/\sigma) L_t(X_i)$ with $L_t = f/g_{m_t, \Sigma_t}$\;
	 Compute the weights $w_i = w'_i / \sum_j w'_j$ with $w'_i = F(\varphi_i / \sigma_{t+1}) L_t(X_i)$\;
	 Compute $m_{t+1} = \sum_{i=1}^N w_i X_i$ and $\Sigma_{t+1} = \sum_{i=1}^N w_i (X_i-m_{t+1})(X_i-m_{t+1})^\top$\;
	 Set $t \longleftarrow t+1$\;
 	 Repeat Steps 2, 3 and 4\;
		}
	Return $\displaystyle \hat P=\dfrac{1}{N} \sum_{i=1}^{N} \II_{\{\varphi(X_i)\geq 0\}}L_t(X_i)$.
	 \caption{Improved CE method of~\cite{papaioannou_improved_2019}.}
	 \label{algo:iCE}
	\end{algorithm}
	

	\newpage
	\bibliographystyle{ieeetr} 
	\bibliography{Bib_Zotero}
\end{document}